# Corpos no interior de um recipiente fechado e transparente em queda livre
*(Free fall of bodies in the interior of a closed and transparent container)*

José Joaquín Lunazzi[*], Leandro Aparecido Nogueira de Paula[**]

Instituto de Física 'Gleb Wataghin'
Universidade Estadual de Campinas-UNICAMP
13083-970 Campinas, São Paulo
*lunazzi@ifi.unicamp.br
**leandroifgw@yahoo.com.br

**Resumo:** Neste artigo discutimos uma nova demonstração experimental da independência das propriedades dos corpos (massa, composição química, forma, etc.) na queda livre. É uma das experiências mais simples, porém uma das mais importantes da Mecânica, tendo sido realizada e repensada repetidamente por diversos cientistas tais como Galileu e Newton. Nossa versão é introduzir dentro de uma garrafa fechada e transparente uma pena e uma pedra observando a queda simultânea destes corpos. Por não haver a necessidade de produzir vácuo, esta versão pode ser repetida por qualquer aluno e professor de ensino médio e universitário em qualquer ambiente, evidenciando sua viabilidade e aplicabilidade na sala de aula.

**Palavras-chave:** vácuo, queda dos corpos, recipiente transparente.

**Abstract:** In this paper we discuss a new experimental demonstration of the independence of the properties of bodies (mass, chemical composition, shape, etc) in free fall. This is one of the simplest experiments in mechanics, though one of the most important ones, having been repeatedly carried out and rethought by several scientists such as Galileo and Newton. Our version of this famous experiment uses one bottle (closed and transparent), in which we introduce a feather and a stone, observing the simultaneous free fall of these bodies. Because it's not necessary to create vacuum, this version can be repeated by teachers and students alike either in secondary schools or universities in any environment. Thus its viability and applicability in the classroom is evident.

Conta-se que Galileu Galilei (1564-1642) após observar o movimento oscilatório do candelabro da catedral em Pisa – indo e vindo – realizou experiências com pêndulos de diferentes massas e comprimentos constatando que o período de oscilação dependia somente do comprimento e não das massas dos pêndulos. Este último resultado levantava sérias suspeitas sobre se a queda dos corpos dependia de suas propriedades (massas, formas, composições químicas, etc.). Então, segundo a lenda (KOYRÉ, 1937), subiu no alto da famosa torre da cidade para lá de cima verificar esta hipótese. Abandonou objetos de pesos diferentes do alto da torre e constatou que eles mantinham suas posições relativas durante a queda, percebendo que a velocidade de queda dos corpos independe de suas massas. Não satisfeito, buscou fazer outra experiência para encontrar a relação entre a distância percorrida e o tempo de queda. Permitindo que esferas de materiais diferentes descessem um plano inclinado, deduziu, então, que a distância percorrida era diretamente proporcional ao quadrado do tempo transcorrido nesta mesma distância (GALILEI, 1988)) e (MORAES; FROTA, 2006). Parecia, então, ter confirmado que a queda dos corpos independe de suas propriedades.

Mas, no entanto, porque no nosso cotidiano ainda temos a impressão que objetos mais leves caem menos rapidamente que objetos mais pesados? Como exemplo, uma experiência simples que pode ser realizada facilmente é o estudo da queda de uma folha de papel e de uma moeda. Comparativamente, quando o papel e a moeda são soltos juntos de uma mesma altura, verificamos que a moeda chega ao chão mais rápido que o papel. O senso comum poderia nos induzir a pensar que esse fato ocorre devido a moeda ser mais pesada que o papel (HÜLSENDEGER, 2004).

Contudo, se amassarmos muito bem o mesmo papel e o soltarmos lado a lado com a moeda, perceberemos agora que o papel amassado e a moeda mantêm suas posições relativas durante a queda caindo praticamente juntos. Amassando o papel não podemos ter conferido a ele mais massa tornando-o mais pesado. Então a sua queda mais rápida deve estar associada à nova forma conferida ao papel ao ser amassado. Mas por que a nova forma implicaria nesta diferença?

Podemos induzir uma resposta ao dizer que a resistência do ar depende da forma do papel. Esta conjetura poderia ser verificada se deixássemos cair o papel e a moeda em uma câmara com bom vácuo ou então ser realizada em ambiente semelhante com uma quantidade muito pequena de fluído. Isaac Newton (1642-1727) ao dizer que *"(...).Pois pequenas penas caindo ao ar livre encontram grande resistência, mas num vidro alto bem esvaziado de ar elas caem tão rápido quanto o chumbo ou o ouro, como me foi dado comprovar diversas vezes(...)"* parece ter verificado esta suposição (NEWTON, 1996). E também os astronautas David Scott e Jim Irwin (1930-1991), pisando em solo lunar pela Apolo 15, realizaram esta experiência ao deixar cair uma pena e um martelo, constatando que caíam praticamente juntas na atmosfera rarefeita da Lua (WOODFILL, 2006).

Uma maneira de tentar demonstrar o mesmo sem necessitar de um tubo sob vácuo vinha sendo realizado por Lunazzi (FERREIRA; LANDERS, 2005) há décadas e consistia em lançar uma moeda ao lado de um pedacinho de papel menor que ela. Após ver a moeda chegar bem antes ao solo, coloca-se o papelzinho por cima da moeda e vê-se como os dois chegam juntos. A hipótese é que a moeda é quem realiza o trabalho de deslocar o ar e portanto o papelzinho cai com sua velocidade natural, a mesma que a moeda teria no vácuo. Pode-se supor até que o papelzinho faça uma pequena pressão na moeda, posto que não experimenta a resistência do ar. O experimento, bem simples, não pode porém ser considerado completo por um motivo ao menos: levando-se em conta que os fenômenos de aerodinâmica são extremamente complexos, nunca podendo ser equacionados corretamente na prática sequer nos casos mais simples pelo fato de que o desejado fluxo laminar nunca acontece, pode-se argumentar que a moeda pode estar arrastando ao papel pelo efeito das correntes de ar que contornam a moeda empurrando o papelzinho contra ela. Embora não saibamos em detalhes o efeito dessas correntes de ar, elas devem existir. Tudo isso pode ser corretamente suposto por um leigo, e se tendemos a não considerá-lo é porque temos o conhecimento preconcebido de que os dois caem com a mesma aceleração.

Em evento semestral da disciplina indicada (FERREIRA; LANDERS, 2005) e (LUNAZZI, 2006), público leigo recebe demonstrações de vários dos experimentos realizados e é assim que acontece uma verdadeira troca de informações, a do professor e alunos que monitoram e as perguntas do público. É nesse trabalho onde pode-se captar o que realmente o experimento pode significar e as muitas abordagens com que deve ser analisado. Foi nesse trabalho que chegamos à evolução do experimento que indicamos neste artigo.

Neste experimento, para sabermos definitivamente se é a resistência do ar a responsável pelo atraso na queda do papel sem amassar, eliminamos durante a queda o efeito direto do ar sobre este mesmo papel. Para isso colocamos equivalentemente, ao invés de papel e moeda, uma pena e uma pedra lado a lado dentro de um recipiente fechado e transparente e observamos a queda, como mostrado na Fig.1 abaixo.

Como neste caso o ar externo não age diretamente sobre os corpos no interior do recipiente, eliminamos sua ação sobre a pena e a pedra, e inclusive sobre a porção de ar dentro do recipiente. Quando abandonamos este sistema de uma determinada altura, o resultado da experiência mostra que a pena e a pedra continuam com suas posições relativas (distância de um em relação ao outro) inalteradas durante a queda.

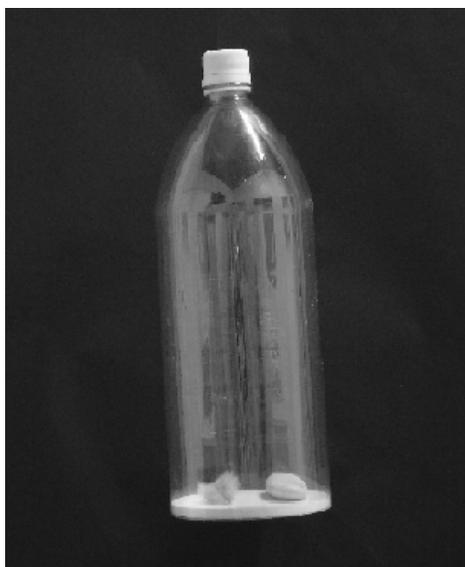

**Fig. 1:** pena (à esquerda) e pedra no interior de um recipiente fechado e transparente simulando uma queda livre. Suas posições relativas (distância de um em relação ao outro) permanecem inalteradas durante a queda.

Uma vez que a pena e a pedra no interior do recipiente possuem massas, formas, composições químicas, etc. distintas, constatamos que a queda dos corpos independe de suas propriedades (massas, formas, composições químicas, etc.) na ausência da resistência do ar e próximo à superfície da Terra.

Contudo, um leigo poderia ainda e tem o direito de pensar que corpos mais leves caem mais lentamente que os mais pesados. E neste caso, a garrafa vista na Fig.1, cairia mais rapidamente que o ar dentro dela. Então, a garrafa, empurraria o ar e este a pena fazendo com que ela caia praticamente junto com a pedra. Para alguém que já aprendeu Física o suficiente a resposta a esta crítica seria clara. Há duas configurações de equilíbrio para o ar dentro da garrafa. Na primeira configuração, quando a garrafa está em repouso com relação a terra, a densidade e pressão de ar são maiores na base diminuindo gradativamente até o topo da garrafa. A segunda configuração de equilíbrio ocorre após a garrafa ter sido abandonada em queda livre. Neste caso, em queda livre, a componente da força gravitacional não mais afetará o conteúdo de ar no interior da garrafa e este tenderá a se homogeneizar. Sua densidade e pressão passarão a ser uniformes em queda livre. Da passagem da primeira para a segunda configuração de equilíbrio, haverá um vento ascendente de ar capaz de deslocar a pena. Este efeito deve ser muito pequeno para ser visto neste experimento, mas sem dúvida físicos experimentais e ou computacionais poderiam tentar detectá-lo. Tendo isso em vista a garrafa não poderia empurrar o ar no seu interior e este a pena. E portanto, a propriedade que queremos demonstrar estaria satisfeita.

Mas devemos tentar responder a crítica do leigo sem o conhecimento preconcebido, baseado em teorias e experimentos já realizados, de como os corpos realmente caem. Pois é a partir deste experimento que tentamos demonstrar a independência das propriedades dos corpos em queda livre. Pois bem, o questionador ainda considera que corpos mais leves caem mais lentamente que os mais pesados e, que o resultado da experiência descrita neste artigo é justamente devido a este fato: a garrafa (mais pesada) cai mais rapidamente que a pena e o ar (mais leves) dentro dela. A garrafa empurra o ar pela parte superior e este, consequentemente, empurra a pena fazendo com que ela caia praticamente junto com a pedra. Mas contra-argumentamos levando em conta apenas uma única propriedade do ar, que também deve ser aceita sem problemas pelo questionador: sua compressibilidade. No entanto ainda manteremos a hipótese de queda dos corpos do questionador e mostraremos que nossa experiência não sustenta esta hipótese.

Visto de dentro da garrafa, a compressibilidade resultaria no início da queda num vento ascendente de ar. Após cessar este vento e haver o equilíbrio, o ar estaria mais denso no topo do que na base . A pena, por ser mais leve que a garrafa, também subiria como no caso do ar e ainda ajudada pelo vento ascendente. Mas na nossa experiência, pelo menos visualmente e de fora da garrafa, não foi verificado que a pena faz este movimento contrário ao movimento da garrafa. A pena permace estática do início ao fim da queda, caindo junto com a pedra. Assim a situação que descrevemos ao considerar a hipótese de queda do leigo e a compressibildade do ar não foi verificada. Eliminamos, ao que parece, todas as possibilidades que poderia fazer com que a pedra e a pena caíssem juntas sem que seja pela  independência das propriedades dos corpos durante a queda. Nos resta então, que esta última esta sendo verificada no nosso experimento. A queda de um corpo não depende da sua massa, forma, composição química, ou qualquer outra propriedade que possa ser atribuída a ele.

Do ponto de vista dos autores, a demonstração experimental apresentada aqui de que a queda dos corpos independe de suas propriedades, na ausência da resistência do ar e próximo à superfície da Terra é um complemento interessante ao experimento do papel amassado e que o reforça usando materiais diferentes, pedra e pena. O leitor interessado poderá facilmente repetir esta experiência.

## Agradecimentos